\documentclass[floats,aps,prb,draft,preprint,showpacs]{revtex4}
\usepackage{psfig}

\begin{document}
\title{Low-temperature conductivity of quasi-one-dimensional conductors: Luttinger
liquid stabilized by impurities}

\author{S.N. Artemenko, S.V. Remizov}

\affiliation{Institute for Radioengineering and Electronics of
Russian Academy of Sciences, 11-7 Mokhovaya str., Moscow 125009,
Russia}

\date{\today}

\begin{abstract}
A new non-Fermi liquid state of quasi-one-dimensional conductors
is suggested in which electronic system exists in a form of
collection of bounded Luttinger liquids stabilized by impurities.
This state is shown to be stable towards inter-chain electron
hopping at low temperatures. Electronic spectrum of the system
contains zero modes and collective excitations of the bounded
Luttinger liquids in the segments between impurities. Zero modes
give rise to randomly distributed localized electronic levels, and
long-range interaction generates the Coulomb gap in the density of
states at the Fermi energy. Mechanism of conductivity at low
temperatures is phonon-assisted hopping via zero-mode states. At
higher voltages the excitations of Luttinger liquid are involved
into electron transport, and conductivity obeys power-law
dependence on voltage. The results are compared with recent
experimental data for NbSe$_3$ and TaS$_3$ crystals.
\end{abstract}

\pacs{71.10.Pm, 71.10.Hf, 71.27.+a, 71.45.Lr}

\maketitle

\section{INTRODUCTION}

The concept of single-electron quasiparticles is one of the
central ideas of solid state physics. Indeed, systems of
interacting electrons in metals and semiconductors are well
described by Landau's Fermi-liquid picture in which interaction
modifies free electrons making them quasiparticles that in many
respects behave like non-interacting electrons. Basic electronic
properties of many solids, including phase transitions into
symmetry-breaking states like superconducting or charge-density
wave (CDW) state, are well described within this concept.
Therefore, possibility to observe non-Fermi liquid behavior in
conducting materials attracts much interest. The concept of
Luttinger liquid is an alternative to the Fermi liquid elaborated
for one-dimensional (1D) electronic systems. It was found that in
1D electronic systems the Fermi-liquid picture breaks down even in
case of arbitrarily weak interaction. Single-electron
quasiparticles cannot exist in 1D metals, and electrons form the
Luttinger liquid in which the only low energy excitations turn out
to be charge and spin collective modes with the sound-like
spectrum. Dynamical independence of these modes gives rise to a
spin-charge separation in 1D systems. Furthermore, correlation
functions at large distances and times decay as a power law with
interaction dependent exponents (for a review see Ref.
\onlinecite{Voit,FishGlaz,SchulzR}).

One may expect to observe formation of Luttinger liquid in
quasi-1D conductors, \textit{i. e.}, highly anisotropic 3D
conductors with chain-like structure. However, according to
present theoretical point of view, the formation of Luttinger
liquid in quasi-1D conductors at low enough temperatures is
problematic because of the instability towards 3D behavior in the
presence of arbitrarily small inter-chain hopping
\cite{PriFirs,Firs,BraYak,Schulz,Boies,Arrig}. So inter-chain
hopping induces a crossover to 3D behavior at low energies, while
Luttinger liquid behavior can survive only at high enough energy
scale where it is not affected by 3D coupling. In contrast to
inter-chain hopping, the Coulomb interaction between electrons on
different chains does not destroy the Luttinger liquid state, it
merely modifies the electronic spectrum of the Luttinger liquid
\cite{Barisic,BoBa,SchulzCoul,MKL}.

Numerous experimental studies of both organic and inorganic
quasi-1D conductors at low temperatures demonstrate, commonly,
Fermi-liquid metallic behavior and/or transitions to
broken-symmetry states, which are described in terms of
Fermi-liquid ideas (for a review see Refs.
\onlinecite{Monceau,Gruner}). For instance, the most studied
inorganic quasi-1D metals (\textit{e. g.}, blue bronze
K$_{0.3}$MoO$_3$, TaS$_3$, (TaSe$_4$)$_2$I \textit{etc}.) undergo
the Peierls transition from metallic to a semiconducting CDW
state, or to semimetallic CDW state like NbSe$_3$. Typically, this
transitions occur in the temperature range $50 \div 250$ K. Some
quasi-1D materials, \textit{e. g.}, TaSe$_3$ do not undergo the
Peierls transition and remain in the normal metallic state at low
temperatures as well.

However, a transition from room-temperature metallic behavior to
nonmetallic one accompanied by disappearance of the CDW state at
temperatures below $50 \div 100$ K was detected in recent experimental
studies of temperature and field dependence of conductivity of
TaS$_3$ and NbSe$_3$ nanoscale-sized crystals
\cite{ZZPM,SHZZZ,ZZ}. The low temperature non-metallic state was
characterized by power law dependencies of conductivity on voltage
and temperature like that expected in Luttinger liquid, or by more
strong temperature dependence corresponding to the variable-range
hopping\cite{ZZ}. Resembling dependencies of conductivity were
reported also in focused-ion beam processed or doped relatively
thick NbSe$_3$ crystals \cite{ZZGSZ}. Transition to non-metallic
conductivity at low temperatures was observed earlier in
fragmented into small pieces and then sintered NbSe$_3$ crystals
\cite{Monceau}. Hopping conductivity was also found out in heavily
doped by iron bulk NbSe$_3$ crystals \cite{Coleman} in dirty
quasi-1D conductors KCP and organic TCNQ-based metals
\cite{organics}, while pure materials are known to undergo the
Peierls transition to the CDW state.

In order to account for such behavior a possibility of
stabilization of the Luttinger state by defects in quasi-1D metals
was put forward recently in the letter of one of the authors
\cite{A}. Such a possibility is based upon finding made in Refs.
\onlinecite{KaneFischer,MatveevGlazman,FuruNag} that a local
impurity potential in Luttinger liquid acts, at low energies, as
effectively infinite barrier. This leads to a discrete spectrum of
collective charge and spin density fluctuations, so that the
electronic system can be considered as a collection of bounded
Luttinger liquids. In this case, at temperatures below the minimum
excitation energy of the collective modes, weak inter-chain
hopping can be considered as small perturbation, and, therefore,
hopping does not destroy the Luttinger liquid state.

However, in order to make comparison with the experimental data
and to find further evidences that the low-temperature
non-metallic state observed in quasi-1D conductors can, actually,
be interpreted as a collection of impurity-induced bounded
Luttinger liquids, calculation of conductivity in the latter state
is needed. Such calculations are the main goal of this paper. We
calculate conductivity  in a collection of weakly coupled metallic
chains with defects in the limit of low temperatures, considering
both the case of the contact interaction used in the standard
Tomonaga-Luttinger model \cite{Voit,SchulzR} and the case of more
physical long-ranged Coulomb potential.

The paper is organized as follows. In Sec. II we present main
equations and formulate the problem in terms of bosonization
technique generalized to the case of long-range Coulomb
interaction between electrons in multi-chain problem in presence
of impurities. In Sec. III we study electronic structure of
quasi-1D conductors. We consider, first, interacting electrons
neglecting inter-chain hopping and find the solution for the phase
operators in the limit of strong impurity potential ignoring, at
the first step, the Coulomb interaction of electrons at different
chains. Modifications induced by the long-range nature of the
Coulomb interaction are discussed in the next subsection. In the
last subsection we show that at low temperatures the inter-chain
hopping can be considered as small perturbation, provided the
hopping matrix element is small enough. Sec. IV is devoted to
transport properties of the system. We show that in the limit of
low voltages conductivity is described in terms of variable-range
hopping, while at large voltages it turns into conductivity
described by a power-law dependence on voltage. And, finally, in
Sec. V we discuss relation of the theoretical results to
experimental data, and make conclusions. In this paper we use
units with Planck and Boltzmann constants equal to unity, $\hbar
=1$, $k_B =1$.

\section{GENERAL EQUATIONS}

First of all we start with the Tomonaga-Luttinger model ignoring
inter-chain hopping integral, $t_\perp$, in order to formulate the
problem in the zero approximation in $t_\perp$. Electronic
operators for right ($r=+1$) and left ($r=-1$) moving electrons
with spin $s$ on chain $n$ are given in terms of phase fields as
(see Ref.\onlinecite{Voit,SchulzR,BoBa})
\begin{eqnarray}
&& \psi_{{\mathbf n},s}(r,x) = \lim_{\alpha\rightarrow
0}\frac{e^{irk_Fx}}{\sqrt{2\pi \alpha}}\eta_{{\mathbf
n},r,s}e^{-iA_{{\mathbf n},r}(x,{\mathbf
n})}, \label{psi} \\
&& A_{{\mathbf n},r} = \frac{1}{\sqrt{2}}[\Theta_{\rho}(x,{\mathbf
n})-r\Phi_{\rho}(x,{\mathbf n})
 + s(\Theta_{\sigma}(x,{\mathbf
n}) - r\Phi_{\sigma}(x,{\mathbf n}))].
 \nonumber
\end{eqnarray}
Phase fields $\Phi_\nu(x,n)$ are related to charge ($\nu=\rho$)
and spin ($\nu=\sigma$) densities, while fields $\Theta_\nu(x)$
are related to the momentum operators $\Pi_\nu = (1/\pi)
\partial_x\Theta_\nu$ canonically conjugate to $\Phi_\nu$. Further,
$\eta_{r,s}$ are Majorana (``real'') Fermionic operators that assure
proper anticommutation relations between electronic operators with
different spin $s$ and chirality $r$, and the cut off length
$\alpha \sim 1/k_F$ is assumed to be of the order of interatomic
distance.

The Hamiltonian of the system includes three terms, the
Hamiltonian of the interacting electrons, the impurity scattering
term, and the inter-chain hopping Hamiltonian. The first term
containing kinetic and potential energies of the electrons is
described by the Tomonaga-Luttinger Hamiltonian generalized for
the multi-chain case with long-range interaction that couples
electrons on different chains \cite{Voit,SchulzR,BoBa}. This
Hamiltonian does ignores the backscattering terms in the
electronic coupling, \textit{i. e.}, it neglects components of
interaction potential with $q_\| \approx 2k_F$. In the bozonized
form the Hamiltonian reads
\begin{equation}
H_0 = \frac{\pi v_{F}d^2}{2} \sum_{\nu=\rho,\sigma} \int
\frac{d{\mathbf q}_\perp dq_\|}{(2 \pi)^3} \left\{ \Pi_\nu^2 +
\frac{1}{\pi^2 K_\nu^2} q_\|^2 \Phi_\nu ^2\right\} \label{H0}
\end{equation}
where $d^2$ is the area per single chain, and integration over
${\mathbf q}_\perp$ is taken within the first Brillouin zone
\begin{equation} \label{Kq}
K_\nu({\mathbf q}_\perp, q_\|)=\left(1+ \frac{2g_\nu}{\pi
v_F}\right)^{-1/2},
\end{equation}
$K_\nu$ is the standard Luttinger liquid parameter describing the
strength of the interaction. $K_\nu$ determines velocities of the
charge ($\nu=\rho$) and spin ($\nu=\sigma$) modes, $v_{\nu} =
v_F/K_\nu$. We study the case of spin-independent interaction,
therefore, in our case, the coupling constant for spin channel is
equal to zero, $g_\sigma =0,$ and $K_\sigma=1$. For charge channel
$g_\rho (\mathbf{q})$ coincides with the matrix element measuring
the strength of the forward scattering due to interaction between
electrons on the same or on different branches of the electronic
spectrum. The exact form of $g_\rho (\mathbf{q})$ depends on
crystalline and electronic structure of a crystal, but in the
long-wavelength limit, $qd \ll 1$, one can describe interaction by
the standard Fourier transform of the Coulomb interaction
potential.
\begin{equation}\label{g}
g_\rho=\frac{4\pi e^2}{d^2(q_\perp^2+q_\|^2)}, \quad v_{\rho}^2 =
v_F^2  + \frac{\omega_{pl}^2}{q_\perp^2+q_\|^2}, \quad
\omega_\rho^2 = q_\|^2v_{\rho}^2,
\end{equation}
where $\omega_{pl}$ is the plasma frequency and $\omega_\rho$ is the 
frequency of the charge mode.

In the coordinate representation
\begin{equation}\label{Kx}
 K_{\rho}^{-2}(x,
{\mathbf n})=\delta_{{\mathbf n}0}\delta (x) +\frac{2e^2}{\pi
v_F\sqrt{x^2+n^2d^2}},
\end{equation}
where the last term is induced by the Coulomb interaction. Its
contribution to the Hamiltonian (\ref{H0}) in the coordinate
representation reads
\begin{equation}
\sum_{\nu=\rho,\sigma} \sum_{\mathbf{n},\mathbf{n}'}\int dx dx'
\left\{\partial_x \Phi_\nu (x,{\mathbf n}) \frac{e^2}{\pi^2
\sqrt{(x-x')^2+(\textbf{n}-\textbf{n}')^2d^2}}
 \partial_{x'} \Phi_\nu (x',{\mathbf n}') \right\}
\end{equation}
Since the operator of the particle density is given by expression
$\rho = -(\sqrt{2}/\pi)\partial_x\Phi(x)$, this term has rather
transparent physical meaning.

The second part of the total Hamiltonian is $2k_F$ impurity
backscattering term \cite{Voit,FishGlaz,SchulzR}
\begin{equation}
H_{imp}=-\sum_{\mathbf{n}}\int dx\sum_i\frac{V(x-x_i)}{\pi \alpha}
\cos{(\sqrt{2} \Phi_\rho + 2k_Fx)}\cos{(\sqrt{2} \Phi_\sigma(x))},
\label{Himp}
\end{equation}
where $V(x)$ is potential of impurity centered in position $i$. In
calculations below we assume $V(x) \approx V_0 \lambda \delta
(x)$, where $V_0$ and $\lambda \sim \alpha$ are amplitude and
radius of the scattering potential, respectively.

When inter-chain electron hopping is ignored, equation of motion
for Heisenberg phase operators $\Phi_\rho$ and $\Theta_\rho$ can
be obtained in standard way from the Hamiltonian $H_0 + H_{imp}$
given by Eqs. (\ref{H0}) and (\ref{Himp}):
\begin{eqnarray}
&& \partial_t  \Theta_\rho(t,x,{\mathbf n}) = v_F \sum_{\mathbf
n'} \int dx' K^{-2}_{\nu}(x-x', {\mathbf n}-{\mathbf
n}')\partial_{x'}\Phi_\nu^{(0)}(t,x',{\mathbf n}')t, \label{Eqmo1} \\
&&
\partial^2_t \Phi_\rho(t,x,{\mathbf n}) - v_F^2 \sum_{\mathbf n'}
\int dx'  \partial_x K^{-2}_{\rho}(x-x', {\mathbf n}-{\mathbf
n}')\partial_{x'}\Phi_\rho(t,x',{\mathbf
n}') \nonumber \\
&& = \frac{\sqrt{2}V_0 d v_F}{\alpha} \sum_i \delta (x-x_i)
\sin{(\sqrt{2} \Phi_\rho + 2k_Fx)}\cos{(\sqrt{2} \Phi_\sigma)}.
\label{Eqmo2}
\end{eqnarray}

Kane and Fischer \cite{KaneFischer} found that the backscattering
impurity potential for repulsive potential ($K_\rho <1$) flows to
infinity under scaling. This conclusion was made for the case of
single chain and a contact interaction potential.

The results of Refs.
\onlinecite{KaneFischer,MatveevGlazman,FuruNag} were generalized
by Fabrizio and Gogolin \cite{FabGogolin} to the case of many
impurities. It was shown that the impurity potential can be
considered as effectively infinite provided that the mean
distance, $\overline{l}$, between impurities satisfies the
condition
\begin{equation}\label{co}
\overline{l} \gg \frac{1}{k_F}
\left(\frac{D}{V_0}\right)^{2/(1-K_\rho)},
\end{equation}
where $D$ is the bandwidth. We assume that the impurity potential
is of atomic scale, $V_0 \lesssim D$, and the interaction between
electrons is not too weak, (\textit{i. e.}, $K_\rho$ is not too
close to 1). Then condition (\ref{co}) is satisfied for
$\overline{l} \gg 1/k_F$, so the limit of strong impurity
potential should be a good approximation in a wide range of
impurity concentrations.

Similar conclusions can be derived for the multiple-chain system
when interaction between electrons on different chains is taken
into account. Indeed, calculations in Ref.
\onlinecite{KaneFischer} are based upon dominant contribution to
Euclidian action expressed in terms of the value of the field
$\Phi_\rho$ at the impurity site by means of solution of equation
of motion in imaginary time. One can act, similarly, in case of
multiple-chain system. We express the solution of  imaginary-time
version of Eq. (\ref{Eqmo2}) for Fourier component of $\Phi_\rho$
in terms of its value at the impurity site, $\Phi_0$, and find
\begin{equation}\label{Ph2}
\Phi (\omega_n,{\mathbf q}) = \frac{\Phi_0}{\omega_n^2 +
q_\|^2v_\rho^2}\left( \int \frac{d^2 d{\mathbf q}_\perp dq_\|}{(2
\pi)^3}\frac{1}{\omega_n^2 + q_\|^2v_\rho^2} \right)^{-1}.
\end{equation}
In the spirit of renormalization group approach the last factor in
Eq. (\ref{Ph2}) must be calculated in the limit of small energies.
Since at small $\omega$ this integral diverges at small $q_\|$, we
use $v_\rho^2=v_F^2/K_\rho^2$ at $q_\|=0$. Then we obtain
$$
\int \frac{d^2 d{\mathbf q}_\perp dq_\|}{(2
\pi)^3}\frac{1}{\omega_n^2 + q_\|^2v_\rho^2}  =
\frac{\overline{K}}{2v_F \omega}, $$ where
$$ \overline{K} \equiv \langle K_\rho ({\mathbf q}_\perp,
q_\|=0) \rangle_{{\mathbf q}_\perp} \equiv \frac{\int d{\mathbf
q}_\perp K_\rho ({\mathbf q}_\perp, q_\|=0) }{ \int d{\mathbf
q}_\perp },
$$
Then using (\ref{Ph2}) one finds the action
\begin{equation}\label{S}
S = \sum_{i\omega_n}\frac{1}{v_F}\left\langle (\omega_n^2 +
q_\|^2v_\rho^2) \Phi^2 (\omega_n,{\mathbf q})
\right\rangle_{\mathbf q} =\sum_{i\omega} 2 \,
\frac{1}{\overline{K}} \, \omega \Phi_0^2(0).
\end{equation}
This equation is similar to the action in Ref.
\onlinecite{KaneFischer} that was used to derive the
renormalization equation for the interaction parameter, the only
difference being that in case of long-range interaction action
(\ref{S}) contains the averaged value $\overline{K}$ of the
interaction parameter instead of parameter $K_\rho$ itself. Thus
impurity potential flows to infinity under scaling if
$\overline{K} < 1$.

So at energies close to the Fermi energy the limit of strong
impurity potential can be used. In the main approximation this
limit corresponds to boundary conditions at the impurity site
\begin{equation}\label{bco}
\sqrt{2} \Phi_\rho + 2k_Fx_i = n \pi, \sqrt{2} \Phi_\sigma = m
\pi,
\end{equation}
where $n$ and $m$ are integers. Furthermore, $n+m$ must be an even
integer in order to ensure the minimum value of the impurity
Hamiltonian (\ref{Himp}). This connection violates independence of
spin and charge modes similar to violation of spin-charge
separation in bounded Luttinger liquid discussed by Eggert
\textit{et al.} \cite{Egg,Mat}.

The last term to be added to the total Hamiltonian describes
next-neighbor inter-chain hopping in the standard way
\begin{eqnarray}
&& H_\perp = t_\perp \sum_{m,n,r,s} \int dx \psi^+_{r,s,m}(x)
\psi_{r,s,n}(x) + H.C. \nonumber \\
&& = \sum_{m,n,r,s} \int dx \frac{it_\perp
\eta_{r,s,n}\eta_{r,s,m}}{\pi \alpha} \left[ \sin(A_{r,m}-A_{r,n})
 + \sin(A_{r,m}-A_{-r,n} + 2irk_Fx) \right],
   \label{Ht}
\end{eqnarray}
where indices $n$ and $m$ denote the chain numbers related to the
nearest neighbors.

\section{ELECTRONIC STRUCTURE AT LOW TEMPERATURES}

We consider first the case of interacting electrons neglecting
inter-chain hopping and find the solution for the phase operators
in the limit of strong impurity potential. According to discussion
in the previous section this limit should be a good approximation
in a wide range of impurity concentrations. In this approximation
the system breaks up into a set of independent segments. At the
beginning we ignore the long-range Coulomb interaction and
consider the case of contact interaction as in the standard
Tomonaga-Luttinger model, and then discuss modifications induced
by long-range interaction. At the end of the section we will show
that at low temperatures the inter-chain hopping does not produce
qualitative modification of the electronic structure based on the
concept of Luttinger liquid.

We begin with solutions for phase operators in case of contact
interaction, taking into account that $\Theta_\nu$ and $\Phi_\nu$
must obey the commutation relations \cite{Voit,FishGlaz,SchulzR}
ensuring anticommutation of electronic operators (\ref{psi}).
Using then the analogy of $H_0$ in Eq.(\ref{H0}) with the
Hamiltonian of an elastic string strongly pinned at impurity
sites, and taking into account commutation relations, we can write
down solutions for the phase operators in the region between
impurity positions at $x=x_i$ and $x_{i+1}$ as
\begin{eqnarray}
&& \Phi_\nu \! = \! \frac{(\delta\phi_{i} \delta_{\nu\rho} - \pi
\Delta N_{\nu i})\tilde x}{l_i} - \pi \! \sum_{j<i} \Delta N_{\nu
j} + \!
 \sum_{n=1}^\infty \! \sqrt\frac{K_\nu}{n}(b_{n\nu i} e^{-i\omega_{n\nu
i} t} + b^+_{n\nu i}e^{i\omega_{n\nu i} t})
\sin{\frac{\pi n \tilde x}{l_i}},\label{fi0} \\
&& \Theta_\nu = \theta_\nu + \frac{(\delta\phi_{i}- \pi \Delta
N_{\nu i})v_{N\nu}t}{l_i} +  i\sum_{n=1}^\infty
\sqrt\frac{1}{K_\nu n}(b_{n\nu i} e^{-i\omega_{n\nu i} t}-
b^+_{n\nu i} e^{i\omega_{n\nu i} t}) \cos{\frac{\pi n \tilde
x}{l_i}}, \label{th0}
\end{eqnarray}
where $\tilde x = x-x_{i},$ $l_i=x_{i+1}-x_{i}$.

Solutions (\ref{fi0}-\ref{th0}) consist of two parts. The terms
with summation present a general solution with zero boundary
conditions, they describe excitations. Excitation spectra of the
eigenmodes are $\omega_{n\nu i} = n \pi v_\nu/l_i$.

The first terms present the particular solution with boundary
conditions (\ref{bco}), they describe zero modes. In the zero
modes
$$v_{N\nu} \equiv \frac{v_F}{K_\nu^2} \equiv
\frac{v_\nu}{K_\nu}, \quad \Delta N_{\rho i} = \frac{(\Delta
N_{\uparrow i} + \Delta N_{\downarrow i})}{\sqrt{2}}, \quad \Delta
N_{\sigma i} = \frac{(\Delta N_{\uparrow i} - \Delta N_{\downarrow
i})}{\sqrt{2}},$$ $\Delta N_{\uparrow i}$ ($\Delta N_{\downarrow
i}$) is the number of extra electrons with spin up (down) in the
region between impurities number $i$ and $(i+1)$,
$\sqrt{2}\delta\phi_{i}$ is equal to the modulo $2\pi$ residue of
$2k_Fl_i$, and, finally $\theta_{\nu i}$ is the phase canonically
conjugate to $\Delta N_{\nu i}$ obeying commutation relations
$[\theta_{\nu i},\Delta N_{\nu i}]=i$.

Note that inside the segments between the impurities the
expressions for the phase fields between the impurity sites turn
out to be similar to those found for bounded 1D Luttinger liquid
with open boundary conditions at the sample ends
\cite{FabGogolin,Egg,Mat}. The main difference is that Majorana
operators $\eta_{s}$ in Eq.(\ref{psi}) are the same for electrons
moving right and left, and that Eq.(\ref{fi0}) contain the
summation over $j < i$ that insure proper commutation relations
between the electron operators related to different segments. Thus
we conclude that the system breaks up into a set of independent
segments described as bounded Luttinger liquids with main
properties similar to those discussed in Refs.
\onlinecite{FabGogolin,Egg,Mat}. In particular, as long as
eigenvalues of $\Delta N_{\uparrow i}$ and $\Delta N_{\downarrow
i}$ are integers we find that $\Delta N_{\rho i}= n_{\rho
i}/\sqrt{2}$ and  $\Delta N_{\sigma i} = n_{\sigma i}/\sqrt{2}$
are not independent, because $n_{\rho i}+n_{\sigma i}$ must be an
even number. Such a limitation ensures the minimum value of the
impurity Hamiltonian (\ref{Himp}). This also means violation of
spin-charge separation due to zero modes as discussed by Eggert et
al. \cite{Egg,Mat} for bounded Luttinger liquids.

Hamiltonian in the region between impurity positions at $x=x_i$
and $x_{i+1}$ can be presented as
\begin{equation}\label{Hb}
H=\sum_\nu (\epsilon_\nu + \sum_n \omega_{n\nu} b^+_{\nu n}b_{\nu
n} ) + \mbox{constant}
\end{equation}
where contribution of zero modes has a form
\begin{equation}\label{e0}
\epsilon_\rho =\frac{v_{N\rho}(\delta\phi_{i}- \pi \Delta
N_{\rho})^2}{2\pi l_i},\quad \epsilon_\sigma =\frac{v_F(\pi \Delta
N_{\sigma})^2}{2\pi l_i}.
\end{equation}

One can see that energies of the zero-mode states depend on number
of extra electrons at a segment confined by impurities.

From the constitutive relation (\ref{psi}) one can see immediately
that temporal dependence (\ref{th0}) of zero-mode part of $\Theta$
should result in local energy levels:
$$
\psi_{{\mathbf n},r,s}(t) \propto
\exp{\left[\frac{1}{\sqrt{2}}(\Theta_{{\mathbf n},\rho} +
s\Theta_{{\mathbf n},\sigma})\right]} \propto
\exp{\left[i\frac{(\delta\phi_{i}- \pi \Delta N_{\rho
i})v_{N\rho}+ \pi \Delta N_{\sigma
i}v_{N\sigma}}{\sqrt{2}l_i}t\right]}.
 $$
Strictly speaking, the single-electron energy levels can be found
from poles of single-particle Green's functions. Its calculation
is similar to that in Ref. \onlinecite{Mat}. As long as we study
the low-temperature behavior of the system, we need explicit
expressions for Green's function at $T \ll \omega_{0,\nu}$. It is
presented in Appendix.

\subsection{Zero modes}

Now we concentrate on zero modes. Consider the zero-mode state
$(n_\rho,n_{\sigma})$ with extra charge $n_\rho$ and extra spin
$n_{\sigma}s$ in one of the segments. Such a state contains
$(n_\rho+n_{\sigma})/2$ extra electrons with spin $s$ and
$(n_\rho-n_{\sigma})/2$ electrons with spin $-s$. Energy
(\ref{e0}) for formation of the zero-mode state
$(n_\rho,n_{\sigma})$ can be conveniently presented in a form (in
this subsection, for brevity, we drop the index related to the
number of a segment).
\begin{equation}\label{e1}
\epsilon = \epsilon_\rho + \epsilon_\sigma, \quad \epsilon_\rho
=\frac{\omega_{0\rho}(\xi - n_\rho)^2}{4},\quad \epsilon_\sigma
=\frac{\omega_{0\sigma}n_{\sigma}^2}{4}, \quad \omega_{0\nu}=
\frac{\pi v_F}{K_\nu^2 l},
\end{equation}
where we introduced a random factor $\xi =\sqrt{2}\delta \phi
/\pi$, $|\xi| <1$. Since factor $\xi$ is related to $2k_Fl$ and,
hence, depends on positions of impurities confining the segment
under consideration, it is different for different segments.

The low-lying frequencies of the single-particle Green's function
$G^{+-}$ and $G^{-+}$ introduced by Keldysh (see Ref.
\onlinecite{Landau}) describe, respectively, energy levels of
electron and hole states. So according to Eqs. (\ref{A11}) and
(\ref{A12}) the energy levels induced by zero-mode states
$(n_{\rho},n_{\sigma})$ are
\begin{equation}\label{om1}
 \varepsilon_{n_\rho n_{\sigma}}= \frac{1}{2}
[\omega_{0\rho} n_\rho + \omega_{0\sigma}n_{\sigma}s
-\omega_{0\rho}(\xi \pm \xi_0)]
\end{equation}
where signs $\pm$ are related to electron and hole states,
respectively, $\xi_0= \frac{1}{2}\left[1+(K_\rho/K_\sigma)^2
\right]$. (We remind that we consider spin-independent
interaction, so in our case $K_\sigma=1$).

As long as typical values of $\xi$ and $\xi_0$ are of the order
unity, a characteristic energy for formation of a zero-mode state
(\ref{e1}) and characteristic energy scale for position of a
zero-mode level (\ref{om1}) can be estimated as
$$\overline{\omega}_0 = \frac{\pi v_F}{K_\nu^2 \overline{l}}, $$
with $\overline{l}=1/N_{imp}$, where $N_{imp}$ is the impurity
density per single chain. If impurities are distributed randomly,
then probability to find a segment of length $l$ is given by the
Poisson distribution, $w(l) = N_{imp} \exp{(-N_{imp} l)}$. Then,
for a given segment, probability to find a value of frequency
$\omega_{0}$ that is much smaller, than $\overline{\omega}_0$, is
exponentially small. Hence, at low temperatures, $T \ll
v_F/\overline{l}$, only energies of modes with
$(n_\rho,n_{\sigma})$ equal to $(0,0), (1, \pm 1)$, and to $(- 1,
\pm 1)$ can be of the order of temperature. Other frequencies
except mentioned above are at least by $\omega_{0}$ larger.
Furthermore, for a given value of $\xi$ not more then two states
can have energies near the Fermi level simultaneously.

So in the statistically averaged zero-mode part of the Green's
function presented in Appendix we can keep only terms related to
the states mentioned above:
\begin{eqnarray}
&& \langle e^{ \frac{i(\delta\phi- \pi \Delta N_{\nu
})(x_a-v_{N\nu}t_a)}{\sqrt{2} l} }\rangle e^{ \frac{\pm
i\pi(x_a-v_{N\nu}t_a)}{4 l} } = \frac{1}{2}
(e^{i(q_{1,1}x_a-\varepsilon_{1,1}
t_a)}+e^{i(q_{1,-1} x_a-\varepsilon_{1,-1}t_a)})n_1 + \nonumber \\
&& \frac{1}{2}(e^{i(q_{-1,1}x_a-\varepsilon_{-1,1}t_a)} +
e^{i(q_{-1,-1} x_a -\varepsilon_{-1,-1}t_a)})n_{-1} + e^{i(q_{0,0}
x_a-\varepsilon_{0,0} t_a)}(1-n_1-n_{-1}), \label{G0}
\end{eqnarray}
with $q_{n_\rho,n_\sigma}= \frac{\pi}{2l}(n_\rho + s n_\sigma \mp
1-\xi)$. Distribution functions for electrons at zero-mode level
are given by expressions
\begin{equation}
n_{\pm 1} = \frac{1}{1+ \frac{1}{2}
\exp{\left[\frac{\omega_{0\rho}(\xi_0\mp\xi)}{2T}\right]}}
\label{n}.
\end{equation}
Note that the distribution functions must not coincide necessarily
with Fermi-Dirac ones because electrons are correlated, and factor
1/2 in the denominator of Eq. (\ref{n}) reflects presence of two
spin states.

Thus zero modes form the system of local electronic levels near
the Fermi energy. The energies of these levels are determined by
random factors $|\xi| <1$, and density of these states at the
Fermi energy is finite in case of short-range interaction.

\subsection{Modifications of zero modes by long-range
interaction}

Zero modes in case of long-range Coulomb interaction can be found
from equations of motion (\ref{Eqmo1}-\ref{Eqmo2}) with time
derivative equal to zero,
\begin{eqnarray}
&& \partial_x \Theta_\nu^{(0)}(x,{\mathbf n}) = 0,\label{Eqz1} \\
&& \Theta_\nu^{(0)}(x,{\mathbf n}) = \theta_\nu +  \sum_{\mathbf
n'} \int dx' v_F K^{-2}_{\nu}(x-x', {\mathbf n}-{\mathbf
n}')\partial_{x'}\Phi_\nu^{(0)}(t,x',{\mathbf n}')t, \label{Eqz2}
\end{eqnarray}
with boundary conditions $\sqrt{2} \Phi_\rho + 2k_Fx = \pi k$,
where $k$ is an integer.

Energy of zero modes in coordinate representation reads
\begin{eqnarray}
&& E = \frac{1}{2\pi} \sum_{\nu,{\mathbf n},{\mathbf n'}} \int dx
\int dx'
 v_F K_{\nu}^{-2}(x-x', {\mathbf n}-{\mathbf n}') \partial_x \Phi_\nu^{(0)}(x,{\mathbf n})
\partial_{x'}\Phi_\nu^{(0)}(x',{\mathbf n}') \nonumber \\
&& = \frac{1}{4} \sum_{\nu,{\mathbf n},{\mathbf n'}} \int dx \int
dx'
 v_F K_{\nu}^{-2}(x-x', {\mathbf n}-{\mathbf n}') \rho_\nu^{(0)}(x,{\mathbf n})
\rho_\nu^{(0)}(x',{\mathbf n}')\label{En}
\end{eqnarray}
with $K_{\nu}$ is determined by Eq. (\ref{Kx}).

Analytical solution of Eqs. (\ref{Eqz1}-\ref{Eqz2}) is not a
simple matter. Therefore, we consider a simplified model in which
we neglect variations of the interaction when each coordinate
varies inside a segment between the impurities. That is, we
approximate interaction parameter $K_{\nu}^{-2}(x-x', {\mathbf
n}-{\mathbf n}')$ by its value, $K_{\nu}^{-2}( {\mathbf
i}-{\mathbf j})$, spatially averaged with respect to coordinates
$x,{\mathbf n}$ and $x',{\mathbf n}'$ inside given segments
labelled as ${\mathbf i}$ and ${\mathbf j}$. This model does not
violate effects of long-range interaction which determine
conductivity at low temperatures calculated in the next section.
It is not difficult to find solution of Eqs.
(\ref{Eqz1}-\ref{Eqz2}) for such model interaction. The solution
for the zero-mode part of the phase operators $\Theta_\nu$ has a
form
\begin{equation}
\Theta_{\nu {\mathbf i}}^{(0)} = \theta_\nu + v_F \sum_{\mathbf j}
K_{\nu}^{-2}( {\mathbf i}-{\mathbf j})(\delta\phi_{\mathbf j}
\delta_{\nu\rho} - \pi \Delta N_{\nu {\mathbf j}})t.\label{ThC}
\end{equation}
This expression for the phase operator enables us to calculate the
energy levels of zero-mode states given by eigenfrequencies of
Green's function. This is not difficult because contributions from
zero modes  are separated from excitation in the Green's function
similar to the case of the short range interaction considered in
the Appendix. So for the energy levels in segment ${\mathbf i}$ we
find
\begin{equation}
\varepsilon_{\mathbf i} = - \frac{v_F}{\sqrt{2}} \sum_{\nu
{\mathbf j}} K_{\nu}^{-2}( {\mathbf i}-{\mathbf
j})(\delta\phi_{\mathbf j}\delta_{\nu\rho} - \pi \Delta N_{\nu
{\mathbf j}}) \pm \frac{\pi v_F}{4}  \sum_{\nu} K_{\nu}^{-2}(0)
\label{eC}
\end{equation}
where signs $\pm$ are, again, related to electron and hole states,
respectively.

For ${\mathbf i} = {\mathbf j}$ the averaged interaction parameter
can be estimated as
\begin{equation}
 K_{\rho}^{-2}(0) \approx  \frac{1}{\overline{l}}\left(1
 +\frac{8e^2}{\hbar v_F} \ln \frac{\overline{l}}{d}\right), \label{Kav}
\end{equation}
where we used dimensional units with $\hbar$ for clarity.

The second term in Eq. (\ref{Kav}) for typical values of Fermi
velocity in quasi-1D conductors is quite large. For $v_F \approx 2
\times 10^7$ cm/s, which is typical value for transition metal
trihalcogenides, its value is about 80. This corresponds to the
case of strong interaction and leads to quite large values of
coupling parameters. So we can estimate the typical energy of
zero-mode levels as
$$\overline{\omega}_0 \sim \frac{e^2}{\overline{l}}. $$

Since interaction factor $K_{\rho}$ has the long-range
contribution, the energy of ``single-electron local levels'' in Eq.
(\ref{eC}) are shifted due to interaction with charges in other
segments. Further, because of the slow decay of interaction factor
$K_{\rho}^{-2}$ with distance described by the Coulomb law, the
summation in Eqs. (\ref{ThC},\ref{eC}) may diverge. In particular,
$\langle\Theta_{\nu {\mathbf i}}^2\rangle$ and $\langle
\varepsilon^2_{{\mathbf i}} \rangle$ diverge for random
distribution of ``number of particles'',
$\sqrt{2}(\delta\phi_{\mathbf j}/\pi- \Delta N_{\nu {\mathbf
j}})\equiv \xi_{\mathbf j} - n_{\nu {\mathbf j}}$, at segments
$\mathbf{j}$. The problem can be resolved by correlated
distribution of localized charges in different segments resulting
in the Coulomb gap at the Fermi energy similar to the case of
localized shallow impurity levels in semiconductors
\cite{ESh,Gant}. Indeed, Eq. (\ref{eC}) has some similarity to the
energy of a local impurity level in the potential induced by all
other impurities. Furthermore, energy of zero-mode local levels
has a form somewhat resembling the expression for the
electrostatic energy of localized impurities:
\begin{equation}
 E = \frac{v_F}{2\pi} \sum_{\nu,{\mathbf i},{\mathbf j}} (\delta\phi_{\mathbf i} \delta_{\nu\rho} - \pi
\Delta N_{\nu {\mathbf i}}) K_{\nu}^{-2}( {\mathbf i}-{\mathbf
j})(\delta\phi_{{\mathbf j}} \delta_{\nu\rho} - \pi \Delta N_{\nu
{\mathbf j}}).
  \label{EC}
\end{equation}

So we adopt to our case arguments by Efros and Shklovskii
\cite{ESh} for the Coulomb gap originally used for a system of
shallow impurities in semiconductors, though there are
considerable differences between equations used in Ref.
\onlinecite{ESh} and Eqs. (\ref{ThC}-\ref{EC}). First, the latter
contain operators of number of extra electrons, $\Delta N_{\nu
{\mathbf i}}$. But at low temperatures only a single eigenvalue of
these operators plays a role, because the states with other
egenvalues $n_{\sigma {\mathbf i}}$ correspond to energies much
larger than temperature. So we can consider the operators as
c-numbers. Another difference is the presence of the terms related
to the spin channel. Below we will see that this does not change
the result.

So following Ref. \onlinecite{ESh} we consider process of transfer
of an electron with spin $s$ from segment ${\mathbf i}$ in the
ground state to segment ${\mathbf j}$. An increase of the energy
of the system induced by such transfer calculated from Eq.
(\ref{EC}), with Eq. (\ref{eC}) taken into account, is equal to
\begin{equation}
\Delta E = \varepsilon_{\mathbf j} - \varepsilon_{\mathbf i} -
\frac{e^2}{r_{\bf ij}} + \frac{\pi v_F n_{\sigma {\mathbf i}}(s +
\delta n_{\sigma {\mathbf i}})}{2l_{\mathbf i}} - \frac{\pi
v_Fn_{\sigma {\mathbf j}}(s - \delta n_{\sigma {\mathbf
j}})}{2l_{\mathbf j}}>0, \label{DE}
\end{equation}
where $\delta n_{\sigma {\mathbf i}}$ is the difference of
$n_{\sigma {\mathbf i}}$ values in the state $i$ after and before
the transfer of the electron, and $r_{\bf ij}$ is the distance
between segments ${\mathbf i}$ and ${\mathbf j}$. Since for the
process of transfer considered here, variation of  $n_{\sigma
{\mathbf i}}$ in the initial state due to removal of an electron
with spin $s$ equals $\delta n_{\sigma {\mathbf i}} = -s$, while
addition of the electron to the final state is described by
$\delta n_{\sigma {\mathbf j}} = s$, two last terms in (\ref{DE})
vanish, and we obtain the relation similar to that derived for the
case of impurity levels in semiconductors. One can see that small
energy differences between the zero-mode levels can be found only
for segments situated far from each other,
\begin{equation}\label{r}
r_{\bf ij} > \frac{e^2}{\varepsilon_{\mathbf j} -
\varepsilon_{\mathbf i}}
\end{equation}
As it was shown in Ref. \onlinecite{ESh}, this results in a
minimum in the density of localized states, $g(\varepsilon)
\propto \varepsilon^2$ near Fermi energy, \textit{i. e.},  to the
Coulomb gap. In particular, for electron transitions with energies
$\Delta E \ll v_F/\overline{l}$ important at low temperatures, the
distance between the segments participating in electron
transitions must be much larger, than the typical segment length.
For such remote segments our simplified model, in which we
neglected the spatial variation of the interaction within distance
of the order of segment length, is asymptotically exact.

\subsection{Effect of long-range interaction on excitation
spectrum}

Solution of equation of motion (\ref{Eqmo2}) with long-range
interaction (\ref{Kx}) satisfying boundary conditions (\ref{bco})
is similar to the problem of interacting segments of elastic
strings of random length. Analytical solution of this problem is
difficult. So, again, we consider a simplified model in which the
impurities are nearly arranged in planes perpendicular to the
conducting chains, so that positions of the impurities differ by
the value much smaller, than the mean distance between the
impurities. Solution of Eqs. (\ref{Eqmo1},\ref{Eqmo2}) consists of
two parts. The first part corresponding to particular solution
with boundary conditions (\ref{bco}) describes zero modes, their
spectrum having the form of the randomly distributed localized
levels discussed above. Levels are situated randomly because the
values of $2k_Fx_i$ which determine positions of zero-mode levels
are random.

The part describing the excitations corresponds to general
solution with zero boundary conditions. The latter can be found by
means of Fourier transformation with respect to chain numbers. We
take into account that according to Eq.(\ref{g}) only excitations
with long wavelengths along the chains correspond to low lying
excitations (otherwise frequencies are of the order of plasma
frequency). So we consider the limit $ q_\perp \gg q_\|$ and
neglect $q_\|$ in the denominator of the interaction parameter
$K_\rho$ (\ref{Kx}). Then we easily find expressions similar to
the last terms in Eqs. (\ref{fi0}-\ref{th0}) describing the
excitations, but with interaction factors and eigenfrequencies
depending on $ q_\perp$,
\begin{equation}\label{omK-C}
\frac{1}{K_{\rho 0}^2}=1+ \frac{8 e^2}{v_F d^2 q_\perp^2},\quad
\omega = n \frac{\pi}{l_i} \sqrt{v_F^2 +
\frac{\omega_{pl}^2}{q_\perp^2}}\,.
\end{equation}
Thus the spectrum of the modes in this case consists of the bands,
the minimum excitation energy being fixed by the maximum value of
$q_\perp$ determined by the transverse reciprocal lattice vector
$\sim \pi/d$. Its characteristic value can be estimated as
\begin{equation}\label{om-est}
 \omega_{min} =  \frac{v_F}{\overline{l}} \sqrt{\pi^2 +
\kappa^2 d^2} = \frac{v_F}{\overline{l}} \sqrt{\pi^2 +
\frac{8e^2}{v_F}},
\end{equation}
where $\kappa =  \omega_{pl}/v_F$ is Thomas-Fermi screening length
in the metallic state.

Contribution of the excitations to the Green's functions can be
calculated similarly to the case of the contact interaction
presented in the Appendix. Factors in the square brackets in
functions $B$ in Eq. (\ref{A2}) originate from summations over
eigenfrequencies in products of phase fields $\Phi_\nu$ and
$\Theta_\nu$. If interaction parameter $K_\rho$ depends on
$q_\perp$ these factors acquire a more complicated form,
\textit{e. g.}, the first factor in square brackets in $B$ must be
substituted for
\begin{equation} \label{F}
\exp \left\langle - \frac{1}{8} \left(\frac{1}{K_{\nu 0}}+ K_{\nu
0} +2 \right) \ln { \left(1-e^{-iz-\tilde \alpha} \right)}
\right\rangle_{{\mathbf q}_\perp},
\end{equation}
where $z = \pi(x_a-v_{\nu} t_a)/l$. Similar substitutions occur in
other factors in square brackets in Eq. (\ref{A2})

In the next section we will need Green's functions for calculation
of conductivity at large voltages, so we need to calculate Green's
functions at energies $\varepsilon \gg \overline{\omega}_0$. Then
in time and coordinate representation we need Green's functions
for $v_{\nu} t_a,  x_a \ll l$. In the same time we consider
$v_{\nu} t_a,  x_a \gg  \alpha$ since the energy is small compared
to Fermi energy. In this limit expression (\ref{F}) reduces to
factor $ \propto z^{-\frac{1}{2}(\delta+1)}$ with
\begin{equation}\label{delta}
\delta = \frac{1}{4}\left\langle \frac{1}{K_{\rho 0}} + K_{\rho 0}
- 2 \right\rangle_{{\mathbf q}_\perp}
\end{equation}
Since density of states $N(\varepsilon)$ is determined by Fourier
transformation of Green's function in $t_a$ at $x_a =0$, and
distribution function, $n(p)$, of electrons on momentum along the
chains can be found by means of Fourier transformation with
respect to $x_a$ at $t_a = 0$, this means that at energies
$\varepsilon \gg \overline{\omega}_0$ $N(\varepsilon) \propto
\varepsilon^{\delta}$ and $n(p) \propto (p-k_F)^{\delta}$,
similarly to the case of Luttinger liquid without impurities
(\textit{cf}. Refs \onlinecite{Voit,SchulzR} and
\onlinecite{BoBa}). Furthermore, if we neglect weak logarithmic
dependence of $v_\rho$ in Eq. (\ref{F}) on ${\mathbf q}_\perp$,
the we find that Greens function in the considered limit behave
like that in the unbounded Luttinger liquid at any relation
between $v_{\nu} t_a$ and $x_a$. In other words, for such
relatively large energies effect of impurities on Green's function
and, hence, on energy spectrum can be neglected.

So we conclude that in the electronic system with Coulomb
interaction the electronic spectrum consists of discrete levels
due to zero modes, and of quasi-continuum spectrum of excitations
of the Luttinger liquid at energies more than
$\overline{\omega}_0$ apart from Fermi energy.

\subsection{Effect of inter-chain hopping}

Now we demonstrate that inter-chain hopping of electrons do not
destroy the description of quasi-1D conductor with impurities at
low temperatures based on the picture of collection of bounded
Luttinger liquid. We take into account the hopping in a standard
way, considering the perturbative series for Green's functions
with respect to the hopping integral $t_\perp$. It is not
difficult to calculate corrections to the Green's functions due to
hopping for small energy $\varepsilon \ll \overline{\omega}_0$,
because at such energies Fourier-transformed Green's functions
presented in the Appendix acquire a simple form. For example, at
such energies the zero-temperature casual Green's function reduces
to a form
\begin{equation}
G^{--} = \frac{\tilde \alpha ^{\frac{1}{4}
(\frac{1}{K_\rho}+K_\rho+2)} [(\tilde \alpha^2+4
\sin^2{}q_0x_1)(\tilde \alpha^2+4 \sin^2{}q_0x_2)]^{\frac{1}{16}
(\frac{1}{K_\rho}-K_\rho)}e^{ik_Fx_a+iq_{\mathbf i} x_a} }{2\pi
\alpha(\varepsilon - \varepsilon_{\mathbf i} + i0 \cdot
\textrm{sign} \varepsilon)}, \label{G00}
\end{equation}
where $\varepsilon_{\mathbf i}$ and $q_{\mathbf i}$ are energy and
wave vector of the zero-level state at segment number ${\mathbf
i}$ (\textit{cf}. Eq. (\ref{G0})).

So calculating in a standard way corrections to the mass operator
in the second order in $t_\perp$ we find that hopping results in a
shift of the zero-mode level by
\begin{equation}
\delta \varepsilon_0  \sim  \sum_{\mathbf j} \frac
{t_\perp^2}{\varepsilon_{\mathbf i} - \varepsilon_{\mathbf j}}
\left( {\frac{\alpha}{l}}\right)^{2\delta}
  \label{t2}
\end{equation}
where $\varepsilon_{\mathbf j}$ are zero-level energies at
neighboring segments.

Note that the probability to find similar values of zero-state
energies at neighboring segments is negligible, so the energy
difference in the denominator of (\ref{t2}) is of the order of
typical energy of the zero-level, $\overline{\omega}_0$. So the
shift of the energy due to hopping can be estimated as
\begin{equation}
\delta \varepsilon_0  \sim   \frac
{t_\perp^2}{\overline{\omega}_0} \left(
{\frac{\alpha}{l}}\right)^{2\delta}. \label{t3}
\end{equation}

The higher order terms give smaller corrections proportional to
higher order powers of factor
 \begin{equation}
\left( \frac {t_\perp}{\overline{\omega}_0}\right)^2 \left(
{\frac{\alpha}{l}}\right)^{2\delta}. \label{t0}
\end{equation}
So corrections are small provided this factor is small.

Note that our calculations are based on discrete nature of the
electron spectrum near the Fermi energy, because we used an
assumption that both characteristic energy and the hopping matrix
element, $t_\perp$, are much smaller than typical energy of
zero-mode levels $\overline{\omega}_0$. The characteristic energy
is determined by temperature, T. Thus inter-chain hopping does not
destroy the Luttinger liquid picture in the limit $T \ll
\overline{\omega}_0$, which does not exist in pure infinite
Luttinger liquid. In the opposite limit of larger temperatures
there are many levels in the energy interval of the order of
temperature, and, hence, the continuum limit is applicable, our
approach is not valid. So at temperatures $T \gg
\overline{\omega}_0 $, the discreteness of the excitation spectrum
cannot be neglected, hence, according to Refs.
\onlinecite{PriFirs,BraYak,Schulz,Boies,Arrig}, inter-chain
hopping is expected to give significant contributions and to
destroy the Luttinger liquid.

\section{Electron transport}

In this section we calculate current in the Luttinger-liquid state
of quasi-1D conductors by means of Keldysh diagrammatic approach
\cite{Landau}.

\subsection{Conductivity at low voltages}

Consider, first, Ohmic conductivity due to hopping between the
segments confined by impurities. Taking into account the
similarity with the case of impurities in semiconductors,
discussed above, it is natural to expect that the mechanism of
conductivity is phonon assisted variable-range hopping.

\begin{figure}[!ht]
  \vskip 0mm
  \centerline{
    \psfig{figure=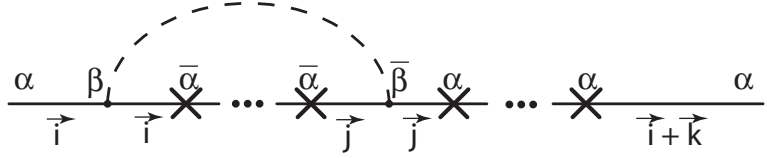,height=3cm
    ,angle=0}
  }
  \caption{A diagram for calculation of the current. The dashed line
symbolizes the phonon propagator, Greek letters denote Keldysh
time indices ``$+$'' and ``$-$'', and $\overline{\alpha}$ means the time
index opposite to $\alpha$. Crosses denote hopping matrix elements
$t_\perp$. Dots between the crosses stand for a way connecting
segments ${\mathbf i}$ and ${\mathbf j}$ corresponding to
sequential transitions via the neighboring segments. The lowest
order contributions to current in $t_\perp$ are given by diagrams
with the least number of transitions.} \label{fig}
\end{figure}

There are two ways for electron transfer: via inter-chain
transitions, and via transitions along conducting chains through
potential barriers induced by impurities. In case of hopping
conductivity there is no principal difference between both
channels. So we concentrate on the case when probability of
transitions between conducting chains dominates over probability
of tunneling through impurities. Current flowing from segment $\mathbf i$
to its neighbors can be calculated as
\begin{equation} \label{I0}
I \propto t_\perp \sum_k [F_{{\mathbf i}, {\mathbf i}+{\mathbf
k}}(1,1) - F_{{\mathbf i}+{\mathbf k},{\mathbf i}}(1,1)],
\end{equation}
where  $F_{{\mathbf i}, {\mathbf i}+{\mathbf k}} =
G^{--}_{{\mathbf i}, {\mathbf i}+{\mathbf k}} + G^{++}_{{\mathbf
i}, {\mathbf i}+{\mathbf k}}$, and summation is performed over
neighboring segments. Green's functions $G^{\alpha
\alpha}_{{\mathbf i}, {\mathbf i}+{\mathbf k}}$ can be calculated
by means of diagrams presented in Fig. \ref{fig}.

Having in mind that conductance is determined by the states with
energies in vicinity of the Fermi energy we can use in the
diagrams electron Green's functions in the form given by Eq.
(\ref{G00}). Performing calculations we find that the total
current through segment ${\mathbf i}$ consists of contributions
describing currents flowing via segments ${\mathbf j}$. Expression
for such current from segment ${\mathbf i}$ to segment ${\mathbf
j}$ induced by voltage difference $V_{\mathbf ij}$ has a form that has
very transparent physical interpretation
 \begin{eqnarray}
&& I_{{\mathbf i}{\mathbf j}} \propto  \int d\omega_{ph} \left[
\frac {t_\perp}{\overline{\omega}_0} \left(
{\frac{\alpha}{\overline{l}}}\right)^{\delta} \right]^{2m}
 \left\{
[n_{\mathbf i}(1-n_{\mathbf j})N_{ph} - n_{\mathbf j}(1-n_{\mathbf
i})(1+N_{ph})]\delta(\varepsilon_{\mathbf i} -
\varepsilon_{\mathbf j}+\omega_{ph}+V_{{\bf ij}})  \right.
\nonumber \\
&& \left. + [n_{\mathbf i}(1-n_{\mathbf j})(1+N_{ph}) - n_{\mathbf
j}(1-n_{\mathbf i})N_{ph}]\delta(\varepsilon_{\mathbf i} -
\varepsilon_{\mathbf j}- \omega_{ph}+ V_{{\bf ij}})\right\}.
\label{Iij}
\end{eqnarray}
Here $\varepsilon_{\mathbf i}, \varepsilon_{\mathbf j}$ and
$n_{\mathbf i}, n_{\mathbf j}$ are electron energies and
distribution functions for zero-mode levels at the respective
segments, and $N_{ph}(\omega_{ph})$ is the distribution function
of phonons. It is clear that different items in Eq. (\ref{Iij})
describe processes of electron hopping between segments
accompanied with phonon emission or absorbtion. The first factor
under the integral originates from virtual transitions between the
segments along the way between segments $\mathbf i$ and $\mathbf j$, and power
index $m$ describes the number of virtual transitions
\begin{equation} \label{m}
m \propto \sqrt{|{\mathbf n}_{\mathbf i} - {\mathbf n}_{\mathbf
j}|^2 + \left(\frac{x_{\mathbf i}-x_{\mathbf
j}}{\overline{l}}\right)^2}.
\end{equation}
Contribution of each virtual transition to the current is
reflected by factor similar to that presented in Eq. (\ref{t0}).

Energies in arguments of the distribution functions in Eq.
(\ref{Iij}) are shifted by values of electro-chemical potential at
given segment. Linearizing expression (\ref{Iij}) with respect to
potentials and performing the integration over phonon frequencies
we find expression for the current between segments $\mathbf i$ and $\mathbf j$
similar to that describing current between localized impurity
states in semiconductors \cite{ESh,Gant}
\begin{equation} \label{I}
 I_{\bf ij} = \frac{V_{\bf ij}}{R_{\bf ij}}, \quad
R_{\bf ij} \propto \exp{\left[2m\,\ln{\left[
\frac{\overline{\omega}_0}{t_\perp}\left(\frac{\overline{l}}{\alpha}
\right)^\delta\right]} + \frac{|\varepsilon_{\mathbf i} -
\varepsilon_{\mathbf j}|}{T}\right]}.
\end{equation}
This expression leads to different results for the case of
short-range interaction of the Tomonaga-Luttinger model, and for
the case of long-range Coulomb interaction of electrons at
different segments. In the former case, following standard
arguments of theory of variable-range hopping \cite{ESh,Gant} we
arrive at Mott's law describing three-dimensional variable-range
hopping conductivity
\begin{equation} \label{s3}
\sigma(T) \propto
\exp{\left[-\left(\frac{T_M}{T}\right)^{1/4}\right]}, \quad T_M
\sim \overline{\omega}_0 \ln^3{\left[
\frac{\overline{\omega}_0}{t_\perp}
\left(\frac{\overline{l}}{\alpha} \right)^\delta\right]}.
\end{equation}

The result is different in case of long-range interaction. Then
there is the Coulomb gap in the density of states at the Fermi
energy, and hopping of electrons occurs effectively only between
the segments situated far from each other, at distances given by
condition (\ref{r})). Since distances between impurities,
$\overline{l}$, is much larger than the interchain distance, $d$, a
given distance $r_{\mathbf i \mathbf j} = \sqrt{|{\mathbf n}_{\mathbf i} - {\mathbf
n}_{\mathbf j}|^2d^2 + (x_{\mathbf i}-x_{\mathbf j})^2}$ can be
found for much smaller values of $m$, (\ref{m}), if the segments
are situated at the same or at adjacent chains, than for segments
situated at remote chains. As current (\ref{I}) between the
segments exponentially decreases with $m$ increasing, the strong
dependence of distance between the segments on direction results
in effectively one-dimensional hopping. Using, again, the
arguments of theory of variable-range hopping we find that
conductivity obeys Efros-Shklovskii law
\begin{equation} \label{sC}
\sigma(T) \propto
\exp{\left[-\left(\frac{T_{ES}}{T}\right)^{1/2}\right]}, \quad 
T_{ES} \sim  \overline{\omega}_0 \ln{\left[
\frac{\overline{\omega}_0}{t_\perp}
\left(\frac{\overline{l}}{\alpha} \right)^\delta\right]}, \quad
\overline{\omega}_0 \sim \frac{e^2}{\overline{l}}.
\end{equation}
Nearly 1D character of hopping results in enhancement of
anisotropy of conductivity. The anisotropy can be estimated as the
ratio of the characteristic hopping lengths in directions parallel
and perpendicular to the chains. It was argued above that hopping
in the perpendicular direction occurs effectively over distances
of the order of inter-chain distance $d$. The hopping length along
the chains can be found as that corresponding to the minimum value
of the exponent in Eq. (\ref{I}). So for anisotropy of the
conductivity we find
\begin{equation} \label{an}
\frac{\sigma_\|}{\sigma_\perp} \sim \frac{e^2}{d \sqrt{T_{ES} T}}.
\end{equation}

\subsection{Non-linear conductivity}

Now we discuss electron transfer at high voltages and make rough
estimates of non-linear current-voltage dependences.

Let the voltage between $m$ segments is equal to $mV$. If the
voltage drop over the hopping length becomes larger then typical
energy of the local levels, $\sim \overline{\omega}_0$, then
electron transitions between the quasi-continuum spectrum of
excitations in segments become possible without phonon absorption.
This happens at electric field $E \gtrsim \sqrt{TT_{ES}} /e
\overline{l}$. Current between such segments can be calculated by
means of diagrams similar to that presented in Fig. \ref{fig}, but
without phonon line. The diagrams contain virtual transitions via
intermediate segments that contribute to the expression for
current  small factors similar to the first factor in Eq.
(\ref{Iij}). Further, at large voltages current is roughly
proportional to factor $(mV)^{2\delta+1}$ because of the power-law
energy dependence of the density of states in Luttinger liquid
(see discussion below Eq. (\ref{delta})).
\begin{equation}
 \label{Im}
I \propto  \left[ \frac {t_\perp}{\overline{\omega}_0} \left(
{\frac{\alpha}{\overline{l}}}\right)^{\delta} \right]^{2m}
(mV)^{2\delta+1}
\end{equation}
Consider, first, the case when the voltage is not too large so
that electron transitions between the regions of quasi-continuum
spectrum of excitations are possible at $m \gg 1$ only. Then index
$m$ is related to the average electric field along the chains by
relation $meV \approx meE \overline{l} \gtrsim
\overline{\omega}_0$. So the most rapid, exponential, dependence
of current on electric field originates from increase of $m$ in
the first factor in Eq. (\ref{Im}), in other word, from switching
on new channels for non-linear current flow between more and more
closely situated segments. The second factor in (\ref{Im}) gives
slowly varying function of field $E$ of the order of
$\overline{\omega}_{0}\,^{2\delta+1}$, its field dependence can be
neglected in comparison to the exponential growth due to the first
factor. So for this regime we find
\begin{equation}
 \label{Ies}
I \propto  \exp{\left( -\frac{E_{0\|}}{E}\right)}, \quad E_{0\|}
\sim \frac{T_{ES}}{e\overline{l}}.
\end{equation}
Note that condition $m > 1$ means $E < E_{0\|}$.

For current in direction perpendicular to conducting chains index
$m$ is related to the average electric field by the relation $meE
d \gtrsim \overline{\omega}_0$. Then for this direction the
characteristic field can be estimated as $E_{0\perp} \sim
T_{ES}/ed$. This is much larger value then $E_{0\|}$ which
demonstrates strong anisotropy of the non-linear conductivity.

At larger voltages, when $E \gtrsim E_0$, the voltage drop is
large enough to induce electron transitions between the
neighboring segments. In this case the I-V curves can be estimated
as
\begin{equation}
 \label{Im1}
I \propto V^{2\delta+1}.
\end{equation}
The power index here is different if tunneling along the chains
via impurities is more effective than inter-chain tunneling. In
the latter case the density of states at the ends of a segment is
described by different power index, similarly to bounded Luttinger
liquids \cite{FabGogolin,Egg,Mat}, so index $2\delta$ in Eq.
(\ref{Im1}) must be substituted for $\langle \frac{1}{K_{\rho
0}}-1 \rangle_{\mathbf{q}_\perp}$.

\section{Discussion}

Now we discuss relation of our theoretical results to real
materials and to experimental data.

Our calculations are based on the Luttinger model which does not
take into account backscattering terms, \textit{i. e.}, $2k_F$
Fourier component of the inter-electronic interaction. These terms
are known to lead to two qualitatively different consequences (for
a review see Ref. \onlinecite{Voit}). In case of attractive
spin-independent interaction it results in the spin gap, while the
backscattering terms are irrelevant if the interaction is
repulsive. So, strictly speaking, our results can be applied for
the case of Coulomb repulsion of the electrons at all wave
vectors. On the other hand, the effective inter-electronic
interaction can be considered as containing both repulsive Coulomb
interaction and attraction due to electron-phonon coupling. It can
happen that the Coulomb repulsion dominates at small values of
wave vectors, while attraction induced by electron-phonon coupling
dominates at wave vectors close to $2k_F$. Though such picture
ignores retardation effects in the electron-phonon coupling it is
quite fruitful in description of quasi-1D CDW-conductors
\cite{Gruner}, electron-phonon interaction being described by
means of attractive electron-electron coupling constant at $2k_F$
wave vector. Note that the mean-field expression for the CDW gap
\cite {Gruner} is similar to the expression for the spin gap
obtained within the renormalization group approach with the
cut-off length $\alpha$ in equation for spin gap estimated as
$\alpha \sim 1/k_F$. Therefore, we can conclude that a possible
value of spin gap in such materials as NbSe$_3$ and TaS$_3$ can be
of the order of few hundreds Kelvins. Since the spin gap implies
long-range order in $\Phi_\sigma$ field and expected values of
spin gap are quite large, the fluctuations of this field at low
temperatures must be small. The energy in the spin channel is
minimized by $\sqrt{2} \Phi_\sigma = 2\pi n$, which satisfies
boundary conditions (\ref{bco}) for even values of integer $m$. So
we can expect that in the presence of large spin gap the electron
system at low temperatures can be considered as spinless electrons
in the Luttinger state, the spin degrees of freedom being frozen.
In this case the results of the preceding sections are still
valid, and only minor rather simplifications than modifications in
the course of derivation of results are needed.

Thus we conclude that our results can be applied to quasi-1D
conductors NbSe$_3$ and TaS$_3$ where transition from metallic to
non-metallic behavior was observed. However, detailed quantitative
comparison with experimental data is difficult. The first reason
for this is some contradictory data on temperature dependence of
Ohmic conductivity at low temperatures. In Ref. \onlinecite{ZZ}
the temperature dependence corresponding to the variable-range
hopping described by the Efros-Shklovskii law was observed, while
in Refs. \onlinecite{ZZPM} and \onlinecite{SHZZZ} a power-law
dependence of conductivity on temperature was observed down to
liquid helium temperature. Another difficulty for making detailed
comparison is that the impurity density in the samples is not
known. However, our results agree with general tendencies observed
in these materials. Namely, more impure samples demonstrate
transition from metallic to non-metallic behavior at higher
temperatures, characteristic temperature in Eq. (\ref{Ies})
increasing from few tens Kelvins in more perfect samples to few
hundreds Kelvins in the most dirty samples. In a sample with
characteristic temperature $E_{ES} \approx 80$ K temperature
dependence of Ohmic conductivity transforms from metallic to
non-metallic behavior in the temperature range 100 - 200 K. Then
we conclude that $\overline{\omega}_0 \sim 100$ K, and since
$\overline{\omega}_0 \sim e^2/\overline{l}$, according to our
approach we can estimate the mean segment length as $\overline{l}
\sim 10^{-5} - 10^{-4}$ cm. According to Eq. (\ref{Ies}) this
length corresponds to electric field of the order of $10^2 - 10^3$
V/cm for transition from Ohm's law to power law dependence. This
is in order of magnitude agreement with the experimental data of
Ref. \onlinecite{ZZ}. But to make convincing conclusions more
theoretical and experimental work is needed.

It is important to note that low temperature behavior of both
NbSe$_3$ and TaS$_3$ in Refs. \onlinecite{ZZPM,ZZGSZ,SHZZZ,ZZ} is
very similar, in spite of the fact that relatively pure samples of
these materials behave quite differently: NbSe$_3$ remains
metallic in the CDW state, while TaS$_3$ below the Peierls
transition becomes an insulator with quasiparticle density obeying
the Arrhenius law. Whereas the main properties of quasi-1D
conductors at higher temperatures are well understood
\cite{Monceau,Gruner}, they demonstrate many intriguing properties
at low temperatures which are still not explained convincingly. In
particular, in TaS$_3$ at temperatures below 20 K a behavior
typical for hopping conductivity was observed \cite{Sa,Zh} instead
of Arrhenius law, and anomalous behavior of the dielectric
function interpreted as a new glassy phase \cite{Biljak} was
detected at frequencies $1 \div 10^7$ Hz. One can speculate that
this unusual behavior can be related to the formation of the
Luttinger phase at very low temperatures because characteristic
energy $\overline{\omega}_0 \sim e^2/\overline{l}$ is small due to
small impurity density $N_{imp} = 1/\overline{l}$.

\section*{ACKNOWLEDGMENTS}

We are grateful and S. V. Zaitsev-Zotov for useful discussions and
to P. Monceau for helpful comments. In part the work was supported
by Russian Foundation for Basic Research (RFBR), by INTAS, and by
CRDF. A part of these research was performed in the frame of the
CNRS-RAS-RFBR Associated European Laboratory ``Physical properties
of coherent electronic states in condensed matter'' between CRTBT
and IRE RAS.

\appendix
\section{Single-electron Green's function}

Here we present Green's functions introduced by Keldysh, $G^{+-}$
and $G^{-+}$.
\begin{equation}
 iG^{+-}_{{\mathbf n},s}(1,2)=\langle \psi_{{\mathbf n},s}(1)\psi_{{\mathbf
n},s}^+(2) \rangle, \quad -iG^{-+}(1,2)= \langle \psi_{{\mathbf
n},s}^+(2) \psi_{{\mathbf n},s}(1) \rangle,
\end{equation}
with notations $1=(r_1,x_1,t_1)$, $2=(r_2,x_2,t_2)$.

These functions are proportional to distribution function of holes
and electrons, respectively. Other Green's functions can be easily
derived by applying time ordering to expressions for $G^{+-}$ and
$G^{-+}$. In particular, the casual Green's function can be found
as $G^{--} = G^{+-}$ at $t_1 > t_2$ and $G^{--} = G^{-+}$ at $t_1
< t_2$.

Performing the calculations in the standard way\cite{Voit,SchulzR}
we find
\begin{equation}\label{A11}
G^{+-}(1,2) = -i B(1,2)\left \langle \exp \left( \frac{i(\delta\phi_{i}- \pi
\Delta N_{\nu i})(x_a-v_{N\nu}t_a)}{\sqrt{2} l_i} \right) \right \rangle \exp \left(
\frac{i\pi(x_a-v_{N\nu}t_a)}{4 l_i} \right)
\end{equation}
The function related to distribution function of electrons has a
form
\begin{equation}\label{A12}
G^{-+}(1,2)= i B(2,1)\left \langle \exp \left( \frac{i(\delta\phi_{i}- \pi
\Delta N_{\nu i})(x_a-v_{N\nu}t_a)}{\sqrt{2} l_i} \right) \right \rangle \exp \left(-
\frac{i\pi(x_a-v_{N\nu}t_a)}{4 l_i} \right)
\end{equation}
Where $B(1,2)$ originates from bosonic excitations of the
Luttinger liquid, and the last factors are related to zero modes.
For $r_1 = +1$ and $r_2 =-1$ this function has a form
$$
B(1,2)= \frac{e^{ik_Fx_a}}{2\pi \alpha}\prod_\nu  \tilde \alpha
^{\frac{1}{4} (\frac{1}{K}+K)} \times
$$
$$
\left\{[\tilde \alpha^2+4 \sin^2{}q_0x_1][\tilde \alpha^2+4
\sin^2{}q_0x_2]\right\}^{\frac{1}{16} (\frac{1}{K}-K)}\times
$$
$$
\left[1-e^{i(\omega_{\nu} t_a -q_0x_a)-\tilde \alpha}
\right]^{-\frac{1}{8} (\frac{1}{K}+K +2)}
\left[1-e^{i(\omega_{\nu} t_a -q_0x_s)-\tilde \alpha}
\right]^{-\frac{1}{8} (\frac{1}{K}-K)} \times
$$
\begin{equation}
\left[1-e^{i(\omega_{\nu} t_a +q_0x_a)-\tilde \alpha}
\right]^{-\frac{1}{8} (\frac{1}{K}+K -2)}
\left[1-e^{i(\omega_{\nu} t_a +q_0x_s)-\tilde \alpha}
\right]^{-\frac{1}{8} (\frac{1}{K}-K)} \label{A2}
\end{equation}
where
 $x_a=x_1-x_2, x_s=x_1+x_2$, $t_a =
t_1-t_2$, $q_0 =\pi/l_i$, $\tilde \alpha = \frac{\pi
\alpha}{l_i}$.

Green's functions with other values of indices $r$ describing
branches of left(right) moving electrons can be found from
relations
$$G(r_1=+1,x_1; r_1= -1,x_2) = G(r_1=+1,x_1; r_1= +1,-x_2),$$
$$G(r_1=-1,x_1; r_1= +1,x_2) = G(r_1=+1,-x_1; r_1= +1,x_2),$$
$$G(r_1=-1,x_1; r_1= -1,x_2) = G(r_1=+1,-x_1; r_1= +1,-x_2).$$

\end{document}